# Moment-Matching Polynomials

Adam Klivans
UT-Austin
klivans@cs.utexas.edu

Raghu Meka
DIMACS and IAS
raghu@ias.edu

**Abstract**

We give a new framework for proving the existence of low-degree, polynomial approximators for Boolean functions with respect to broad classes of non-product distributions. Our proofs use techniques related to the classical moment problem and deviate significantly from known Fourier-based methods, which require the underlying distribution to have some product structure.

Our main application is the first polynomial-time algorithm for agnostically learning any function of a constant number of halfspaces with respect to any log-concave distribution (for any constant accuracy parameter). This result was not known even for the case of learning the intersection of two halfspaces without noise. Additionally, we show that in the *smoothed-analysis* setting, the above results hold with respect to distributions that have sub-exponential tails, a property satisfied by many natural and well-studied distributions in machine learning.

Given that our algorithms can be implemented using Support Vector Machines (SVMs) with a polynomial kernel, these results give a rigorous theoretical explanation as to why many kernel methods work so well in practice.

# 1 Introduction: Beyond Worst-Case Learning Models

Learning halfspaces is one of the core algorithmic tasks in machine learning and can be solved in the noiseless (PAC) model via efficient algorithms for linear programming. The two simplest generalizations of this problem namely 1) learning the intersection of two halfspaces and 2) learning a noisy halfspace (i.e., agnostic learning) have attracted the attention of many researchers in theoretical computer science and statistics. Surprisingly, they both remain challenging open problems.

In the context of computational complexity, there are many *hardness* results for learning halfspace-related concept classes with respect to arbitrary distributions, and the literature is too vast for us to survey here. A brief summary might be that strong NP-hardness results are known for *proper* learning, where the learner must output a hypothesis that is of the same form (or close to the same form) as the unknown concept class [FGRW, GR, DOSW, KS1], and that there are cryptographic hardness results even for improper learning, where the learner is allowed to output *any* polynomial-time computable hypothesis [FGKP, KS2]. These hardness results apply to many easy-to-state problems, including the two simple generalizations of learning halfspaces enumerated above.

There is a disconnect, however, between the many discouraging hardness results for learning convex sets and the success in practice of popular machine-learning tools for solving just these types of problems (e.g., Support Vector Machines). A reasonable question might be "Why do kernel methods– algorithms that at their core learn a noisy halfspace– work so well in practice?" The allusion here to Spielman and Teng's work on Smoothed Analysis [ST] is on purpose: supervised learning seems perfectly suited to an average-case analysis in terms of the underlying distribution on examples.

Indeed, the main positive result of this paper is a smoothed analysis of learning functions of halfspaces: we show that, in the smoothed-analysis setting, functions of halfspaces are agnostically learnable with respect to any distribution that obeys a subexponential tail bound (so-called subexponential densities) for any constant error parameter. These distributions include all logconcave distributions and need not be product or unimodal. Previous work (that we detail in the next section) required the underlying distribution to be Gaussian or uniform over the hypercube.

We leave open the possibility that functions of halfspaces are agnostically learnable with respect to *all* distributions in the smoothed-analysis model (i.e., all distributions that have been subject to a small Gaussian perturbation). We certainly are not aware of any, say, cryptographic hardness results for this setting.

## 1.1 Introduction: Previous Work on Distribution-Specific Learning

Many researchers have studied the complexity of learning convex sets with respect to fixed marginal distributions. Along these lines, Blum and Kannan [BK1] gave the first polynomial-time algorithm for learning intersections of $m = O(1)$ halfspaces with respect to Gaussian distributions on $\mathbb{R}^n$. Their algorithm runs in time $n^{O(2^m)}$ (for any constant accuracy parameter). Vempala [Vem2] improved on this work and gave a randomized algorithm for learning intersections of centered halfspaces with respect to any log-concave distribution on $\mathbb{R}^n$ in time roughly $(n/\varepsilon)^{O(m)}$ ("centered" means that each bounding hyperplane passes through the mean of the distribution). In a beautiful follow-up paper, Vempala [Vem1] used PCA to give an algorithm for learning the inter-section of $m$ halfspaces with respect to any Gaussian distribution in time $\mathsf{poly}(n) \cdot (m/\varepsilon)^{O(m)}$. We note that these results hold in the PAC model, and it is not clear if they succeed in the agnostic setting.

In the agnostic model, we are only aware of results that use the polynomial regression algorithm of Kalai et al. [KKMS]. Klivans et al. [KOS1] (combined with the observations in Kalai et al.) gave an algorithm for learning *any function* of $m$ halfspaces in time $n^{O(m^2/\varepsilon^2)}$ with respect to the uniform distribution on $\{-1,1\}^n$. Applying results on Gaussian surface area, Klivans et al. [KOS2] gave an algorithm for agnostically learning intersections of $m$ halfspaces in time $n^{\mathsf{polylog}(m)/\varepsilon^2}$ with respect to any Gaussian distribution.

A major goal in this area has been to move beyond Gaussians and tackle the case when the underlying distribution is log-concave, as log-concave densities are a broad and widely-studied class of distributions. The Gaussian density is log-concave, and, in fact, *any* uniform distribution over a convex set is log-concave.

Kalai et al. [KKMS] give an algorithm for agnostically learning a *single* halfspace with respect to any log-concave distribution in time $n^{f(\varepsilon)}$ for some function $f$. The best known bound for $f$ is currently $2^{O(1/\varepsilon^2)}$ (follows from Section 5 of Lubinsky [Lub]). It is unclear how to extend the Kalai et al. analysis to work for the intersection of two halfspaces.

To summarize, it was not known how to learn the intersection of two halfspaces with respect to log-concave distributions even in the noiseless (PAC) model.

## 1.2 Statement of Results

Here we give the first polynomial-time algorithm for agnostically learning intersections (or even arbitrary functions) of a constant number of halfspaces with respect to any log-concave distribution on $\mathbb{R}^n$(see Table 1.2 for the precise parameters):

**Theorem 1.1.** *Functions of $m$ halfspaces are agnostically learnable with respect to any log-concave distribution on $\mathbb{R}^n$ in time $n^{O_{m,\varepsilon}(1)}$ where $\varepsilon$ is the accuracy parameter.*

Admittedly, our dependence on the number of halfspaces $m$ and the error parameter $\varepsilon$ is not great, but we stress that no polynomial-time algorithm was known even for the intersection of two halfspaces. See Table 1.2 for a summary of previous work.

We remark that Daniel Kane in a forthcoming paper has independently obtained Theorem 1.1 using a set of completely different techniques ([Kan1]). His dependence on $m$ and $1/\varepsilon$, though still exponential, is superior to ours.

We extend the above result– in the *smoothed-analysis* setting– to hold with respect to arbitrary distributions with sub-exponential tail bounds. We first define the model of smoothed-complexity that we consider.

**Definition 1.2.** *Given a distribution $\mathcal{D}$ on $\mathbb{R}^n$, and a parameter $\sigma \in (0,1)$, let $\mathcal{D}(\sigma)$ be a perturbed distribution of $\mathcal{D}$ obtained by independently picking $X \leftarrow \mathcal{D}$, $Z \leftarrow \mathcal{N}(0,\Sigma)^n$ and outputting $X + Z$, where $\Sigma \succeq \sigma \cdot cov(X)$[1].*

That is, $\mathcal{D}(\sigma)$ is obtained by adding Gaussian noise to $\mathcal{D}$ and quantitatively, we want the variance of the noise in any direction to be comparable to (at least $\sigma^2$ times) the variance of $\mathcal{D}$ in the same direction. For instance, for $\mathcal{D}$ isotropic, perturbations by $\mathcal{N}(0,\sigma)^n$ would suffice.

---
[1]Here, $\succeq$ denotes the semi-definite ordering.

| Concept Class | Distribution | Running Time | Model | Source |
|---|---|---|---|---|
| Intersections | Gaussian | $\mathsf{poly}(n) \cdot (m/\varepsilon)^m$ | PAC | [Vem1] |
| Intersections | Gaussian | $n^{(\mathsf{polylog}(m)/\varepsilon^{O(1)})}$ | Agnostic | [KOS2] |
| Intersections (centered) | Log-concave | $(n/\varepsilon)^m$ | PAC | [Vem2] |
| One halfspace | Log-concave | $n^{f(\varepsilon)}$ | Agnostic | [KKMS] |
| Arbitrary | Log-concave | $n^{\exp((\log(1/\varepsilon))^{\tilde{O}(m)}/\varepsilon^4)}$ | Agnostic (moment-matching) | This work |
| Arbitrary | Sub-exponential | $n^{\exp((\log(\log m/\sigma\varepsilon))^{\tilde{O}(m)}/\sigma^4\varepsilon^4)}$ | Agnostic ($\sigma$-smoothed) | This work |
| Arbitrary | Sub-gaussian | $n^{(\log(\log m/\sigma\varepsilon))^{\tilde{O}(m)}/\sigma^4\varepsilon^4}$ | Agnostic ($\sigma$-smoothed) | This work |

Figure 1: Summary of recent work on learning intersections and arbitrary functions of $m$ halfspaces

The latter corresponds more directly to the traditional smoothed-complexity setup, but we use the above definition as it is basis independent and allows for non-spherical Gaussian perturbations.

We define the smoothed-complexity of (agnostically) learning a concept class $\mathcal{C}$ under a distribution $\mathcal{D}$ to be the complexity of (agnostically) learning $\mathcal{C}$ under the perturbed distributions $\mathcal{D}(\sigma)$. This model first appears in the work of Blum and Dunagan [BD] (for the special case of spherical Gaussian perturbations) and we believe it to be a natural and practical extension of the traditional models of learning. For instance, the main motivating principle behind smoothed-analysis– that real data involves measurement error– is very much applicable here. Besides the work of Blum and Dunagan, there seems to be little known about learning in this model.

We say a distribution is sub-exponential (sub-gaussian) if every marginal (i.e., one-dimensional projection) of the distribution obeys a tail bound of the form $e^{-|z|}$ ($e^{-|z|^2}$, respectively). It is known that all log-concave distributions are sub-exponential. Sub-exponential and sub-gaussian densities are commonly studied in machine learning and statistics and model various real-word situations (see [BK2] for instance). We show that for these types of distributions, our learning algorithms have polynomial smoothed-complexity (for constant $\sigma$):

**Theorem 1.3.** *Functions of $m$ halfspaces are agnostically learnable with respect to any sub-exponential distribution on $\mathbb{R}^n$ in time $n^{O_{m,\varepsilon,\sigma}(1)}$ where $\varepsilon$ is the accuracy parameter and $\sigma$ is the perturbation parameter.*

We obtain much better parameters (in the constant hidden in $O_{m,\varepsilon,\sigma}(1)$) for the special case of sub-gaussian densities (see Theorem 4.2).

Blum and Dunagan were the first to study the smoothed complexity of learning halfspaces. They showed that for a single halfspace in the noiseless (in labels) setting, the perceptron algorithm converges quickly with high probability for examples perturbed by Gaussian noise. Their expected running time, however, is infinite (and thus strictly speaking does not give bounds on the smoothed-complexity of the Perceptron algorithm).

To obtain our smoothed-analysis results, we prove that Gaussian perturbations provide enough *anti concentration* for our polynomial approximation methods to work. We believe this connection will find additional applications related to the smoothed-complexity of learning Boolean functions.

## 1.3 Overview of Conceptual and Technical Contributions

In their seminal paper, Linial et al. [LMN] introduced the polynomial approximation approach for learning Boolean functions. The core of their approach is to solve the following optimization problem: given a Boolean function $f$, minimize, over all polynomials $p$ of degree at most $d$, the quantity $\mathbb{E}_{x \in \{-1,1\}^n}[(f-p)^2]$.

The algorithm is given uniformly random samples of the form $(x, f(x))$. Their "low-degree" algorithm approximately solves this optimization problem in time roughly $n^{O(d)}$. Later, the "sparse" algorithm of Kushilevitz and Mansour [KM2] solved the same optimization problem but where the minimization is over all sparse polynomials, and the algorithm is allowed query access to the function $f$. These algorithms were developed in the context of PAC learning.

Kalai et al. [KKMS] subsequently observed that in order to succeed in the agnostic framework of learning (we formally define agnostic learning in Section 2.1 but for now agnostic learning can be thought of as a model of PAC learning with adversarial noise), it suffices to approximately minimize $\mathbb{E}_{x \in \{-1,1\}^n}[|f-p|]$.

That is, minimizing with respect to the 1-norm rather than the 2-norm results in highly noise-tolerant learning algorithms. Finding efficient algorithms for directly minimizing the above expectation with respect to the 1-norm ("$\ell_1$ minimization"), however, is more challenging than in the $\ell_2$ case. The work of Kalai et al. [KKMS] gives the analogue of the "low-degree" algorithm for $\ell_1$ minimization (in fact, their algorithm can be carried out using a Support Vector Machine with the polynomial kernel), and the work of Gopalan et al. [GKK] gives the analogue of the "sparse" algorithm for $\ell_1$ minimization.

Although we have efficient algorithms that directly carry out $\ell_1$ minimization for low-degree polynomials, proving the *existence* of good low-degree $\ell_1$ approximators has required first finding a good low-degree $\ell_2$ approximator (i.e., Fourier polynomial) and then applying the simple fact that $\mathbb{E}[|p|] \leq \sqrt{\mathbb{E}[p^2]}$. Directly analyzing the error of low-degree $\ell_1$ approximators seems quite difficult. In our setting, for example, it is not even clear that the best low-degree $\ell_1$ polynomial approximator is unique!

The main conceptual contribution of our methods is to provide the first framework for *directly* proving the existence of low-degree $\ell_1$ approximating polynomials for Boolean functions (in fact, we also obtain *sandwiching* polynomials). One benefit of our approach is that we do not require the underlying distribution to be product (essentially all of the techniques involving the discrete Fourier polynomial require some sort of product structure). As such, in this work, we are able to reason about approximating Boolean functions with respect to interesting non-product distributions, such as log-concave densities.

In the following descriptions, we assume we are trying to show polynomial approximations for $f : \mathbb{R}^n \to \{0, 1\}$, where $f = g(h_1(x), \ldots, h_m(x))$, where $g : \{0, 1\}^m \to \{0, 1\}$ is an arbitrary Boolean function and $h_1, \ldots, h_m : \mathbb{R}^n \to \{0, 1\}$ are halfspaces.

## 1.4 A "Moment-Matching" Proof

Our method uses ideas from probability theory and linear programming to give a framework for proving the existence of sandwiching polynomials (it is easy to see that sandwiching polynomials are stronger than $\ell_1$ approximators). The main technical contribution is to show how to use a set of powerful theorems from the study of the *classical moment problem* to apply our framework to functions of halfspaces. At a high level, our approach makes crucial use of the following consequence of strong duality for semi-infinite linear programs: let $\mathcal{D}$ be a distribution and let $\mathcal{D}_k$ be any distribution where all moments of order less than or equal to $k$ match those of $\mathcal{D}$. If $\mathbb{E}_\mathcal{D}[f]$ is "close" to $\mathbb{E}_{\mathcal{D}_k}[f]$ then $f$ has a low-degree sandwiching polynomials with respect to $D$. The question then becomes how to analyze the bias of a Boolean function where only the low-order moments of a distribution have been specified. We show how to use several deep results from probability to answer this question in Sections 3.2 and 3.3.

We show that the moment-matching approach also has some interesting applications for learning with respect to distributions on the discrete cube $\{-1, +1\}^n$.

## 2 Preliminaries

### 2.1 Agnostic Learning

We recall the model of agnostically learning a concept class $\mathcal{C}$ [Hau], [KSS]. In this scenario there is an unknown distribution $\mathcal{D}$ over $\mathbb{R}^n \times \{-1, 1\}$ with marginal distribution over $\mathbb{R}^n$ denoted $\mathcal{D}_X$. Let $\mathsf{opt} \stackrel{\text{def}}{=} \inf_{f \in \mathcal{C}} \Pr_{(x,y) \sim \mathcal{D}}[f(x) \neq y]$; i.e. opt is the minimum error of any function from $\mathcal{C}$ in predicting the labels $y$. The learner must output a hypothesis whose error is within $\varepsilon$ of opt:

**Definition 2.1.** *Let $\mathcal{D}$ be an arbitrary distribution on $\mathbb{R}^n \times \{-1, 1\}$ whose marginal over $\mathbb{R}^n$ is $\mathcal{D}_X$, and let $\mathcal{C}$ be a class of Boolean functions $f : \mathbb{R}^n \to \{-1, 1\}$. We say that algorithm $B$ is an agnostic learning algorithm for $\mathcal{C}$ with respect to $\mathcal{D}$ if the following holds: for any $\mathcal{D}$ as described above, if $B$ is given access to a set of labeled examples $(x, y)$ drawn from $\mathcal{D}$, then with probability at least $1 - \delta$ algorithm $B$ outputs a hypothesis $h : \mathbb{R}^n \to \{-1, 1\}$ such that $\Pr_{(x,y) \sim \mathcal{D}}[h(x) \neq y] \leq \mathsf{opt} + \varepsilon$.*

Note that PAC learning is a special case of agnostic learning (the case when $\mathsf{opt} = 0$).

The "$L_1$ Polynomial Regression Algorithm" due to Kalai et al. [KKMS] shows that one can *agnostically* learn any concept class that can be approximated by low-degree polynomials (in Kalai et al. [KKMS] it is shown how to implement this algorithm using a standard SVM with the polynomial kernel) :

**Theorem 2.2** ([KKMS]). *Fix $\mathcal{D}$ on $X \times \mathbb{R}$ and let $f \in \mathcal{C}$. Assume there exists a polynomial $p$ of degree $d$ such that $\mathbb{E}_{x \sim \mathcal{D}_X}[|f(x) - p(x)|] < \varepsilon$ where $\mathcal{D}_X$ is the marginal distribution on $X$. Then, with probability $1 - \delta$, the $L_1$ Polynomial Regression Algorithm outputs a hypothesis $h$ such that $Pr_{(x,y) \sim \mathcal{D}}[h(x) \neq y] \leq opt + \varepsilon$ in time $\mathrm{poly}(n^d/\varepsilon, \log(1/\delta))$.*

Throughout, we suppress the $\mathrm{poly}(\log(1/\delta))$ dependence on $\delta$.

## 2.2 Probability

For a random variable $X \in \mathbb{R}^m$, let $\varphi_X : \mathbb{R}^m \to \mathbb{R}$ be the characteristic function defined by $\varphi_X(t) = \mathbb{E}[\exp(-i\langle t, x\rangle)]$, where $i = \sqrt{-1}$.

We shall use the following standard distance measures between random variables $X, Y \in \mathbb{R}^m$.

- The $\lambda$-metric:
$$\mathsf{d}_\lambda(X, Y) = \min_{T>0} \max\{\max_{\|t\|\leq T}\{|\varphi_X(t) - \varphi_Y(t)|\}, 1/T\}.$$

- The Levy distance: for $\mathbf{1}$ being the all 1's vector,
$$\mathsf{d}_{\mathsf{LV}}(X, Y) = \inf_{\varepsilon>0} \{\forall t \in \mathbb{R}^m, \ \Pr[X < t - \varepsilon\mathbf{1}] - \varepsilon < \Pr[Y < t] < \Pr[X < t + \varepsilon\mathbf{1}] + \varepsilon\}.$$

- Kolmogorov-Smirnov or cdf distance:
$$\mathsf{d}_{\mathsf{cdf}}(X, Y) = \sup_{t \in \mathbb{R}^m} \{|\Pr[X \geq t] - \Pr[Y \geq t]|\}.$$

For $I = (i_1, \ldots, i_n) \in \mathbb{Z}^n$, and $x \in \mathbb{R}^n$, let $x(I) = \prod_{j=1}^n x_j^{i_j}$. For $k > 0$, let $I(k, n) = \{I = (i_1, \ldots, i_n) \in \mathbb{Z}^n : \sum_{j=1}^n i_j \leq k, \ i_j \geq 0\}$.

We say that a class of functions $\mathcal{C}$ is $\varepsilon$-approximated in $\ell_1$ by polynomials of degree $d$ under a distribution $\mathcal{D}$ if for every $f \in \mathcal{C}$, there exists a degree $d$ polynomial $p$ such that $E_{x \sim \mathcal{D}}[|p(x) - f(x)|] \leq \varepsilon$.

We use the following properties of log-concave distributions (equivalent formulations can be found in Lovasz-Vempala [LV]).

**Theorem 2.3** ([CW]). *Let random-variable $X \in \mathbb{R}^n$ be drawn from a log-concave distribution. Then, for every $w \in \mathbb{R}^n$, and $r > 0$, $\mathbb{E}[|\langle w, X\rangle|^r] \leq r^r \cdot \mathbb{E}[\langle w, X\rangle^2]^{r/2}$.*

**Theorem 2.4** ([CW]). *There exists a universal constant $C$ such that the following holds. For any real-valued log-concave random variable $X$ with $\mathbb{E}[X^2] = 1$ and all $t \in \mathbb{R}$, $\varepsilon > 0$, $\Pr[X \in [t, t+\varepsilon]] < C\varepsilon$.*

We also use the following simple lemmas. The first helps us convert closeness in Levy distance to closeness in cdf distance, while the second helps us go from fooling intersections of halfspaces to fooling arbitrary functions of halfspaces.

**Fact 2.5.** *Let $X = (X_1, \ldots, X_m) \in \mathbb{R}^m$ be a random variable such that for every $r \in [m]$, $t \in \mathbb{R}, \varepsilon > 0$, $\Pr[X_r \in [t, t+\varepsilon]] < \beta \cdot \varepsilon$ for a fixed $\beta > 0$. Then, for any random variable $Y$, $\mathsf{d}_{\mathsf{cdf}}(X, Y) \leq m \cdot \beta \cdot \mathsf{d}_{\mathsf{LV}}(X, Y)$.*

**Lemma 2.6.** *Let $X, Y \in \mathbb{R}^m$ be real-valued random variables such that for every $a_1, \ldots, a_m \in \{1, -1\}$, $\mathsf{d}_{\mathsf{cdf}}((a_1 X_1, a_2 X_2, \ldots, a_m X_m), (a_1 Y_1, a_2 Y_2, \ldots, a_m Y_m)) \leq \varepsilon$. Then, for any function $g : \{1, -1\}^m \to \{1, -1\}$ and thresholds $\theta_1, \ldots, \theta_m$, $|\mathbb{E}[g(\mathsf{sign}(X_1 - \theta_1), \ldots, \mathsf{sign}(X_m - \theta_m))] - \mathbb{E}[g(\mathsf{sign}(Y_1 - \theta_1), \ldots, \mathsf{sign}(Y_m - \theta_m))]| \leq 2^m \varepsilon$.*

*Proof.* Fix $\theta_1, \ldots, \theta_m$ and let $X' = (\mathsf{sign}(X_1 - \theta_1), \ldots, \mathsf{sign}(X_m - \theta_m))$ and define $Y'$ similarly. Then, from the assumptions of the lemma, for every $a \in \{1, -1\}^m$,

$$|\Pr[X' = a] - \Pr[Y' = a]| < \mathsf{d}_{\mathsf{cdf}}((a_1 X_1, a_2 X_2, \ldots, a_m X_m), (a_1 Y_1, a_2 Y_2, \ldots, a_m Y_m)) < \varepsilon.$$

Therefore, $\mathsf{d}_{\mathsf{TV}}(X', Y') < 2^{m-1}\varepsilon$. The lemma now follows. $\square$

# 3 Moment-Matching Polynomials

We develop a theory of "moment-matching polynomials" for showing the existence of good approximating polynomials. Our main result is the following.

**Theorem 3.1.** *Let $\mathcal{D}$ be a log-concave distribution over $\mathbb{R}^n$. Let $h_1, \ldots, h_m : \mathbb{R}^n \to \{1, -1\}$, be halfspaces and let $g : \{1, -1\}^m \to \{1, -1\}$ be an arbitrary function. Define $f : \mathbb{R}^n \to \{1, -1\}$ by $f(x) = g((h_1(x), \ldots, h_m(x)))$. Then, there exists a real-valued polynomial $P$ of degree at most $k = \exp((\log((\log m)/\varepsilon))^{O(m)}/\varepsilon^4)$ such that $\mathbb{E}_{X \leftarrow \mathcal{D}}[|f(X) - P(X)|] \leq \varepsilon$.*

Theorem 1.1 with the runtime given in Table 1.2 follows from the above result and Theorem 2.2. The theorem is proved in Section 3.3. We start by describing the two basic ingredients: LP-duality and the classical moment problem.

## 3.1 LP Duality

It is now well known in the pseudorandomness literature that with respect to the uniform distribution over $\{-1, 1\}^n$, a concept class $\mathcal{C}$ has degree $k$ sandwiching polynomials if and only if $\mathcal{C}$ is fooled by $k$-wise independent distributions [Baz]. The proof of this fact follows from LP duality where feasible solutions to the primal are $k$-wise independent distributions and feasible dual solutions are approximating polynomials.

In our setting, we consider continuous distributions over $\mathbb{R}^n$ that are not necessarily product. As such, this equivalence is more subtle. In fact, it is not even clear how to define $k$-wise independence for non-product distributions (such as log-concave densities). Still, given a distribution $\mathcal{D}$ we can write a semi-infinite linear program (a program with infinitely many variables but finitely many constraints) whose feasible solutions are distributions that match all of $\mathcal{D}$'s moments up to degree $k$ (in the case where $\mathcal{D}$ is uniform over $\{-1, 1\}^n$, matching all moments is equivalent to being $k$-wise independent).

For $I \in I(k, n)$, let $\sigma_I = \mathbb{E}_{X \leftarrow \mathcal{D}}[X(I)]$. Let $f \in \mathcal{C}$. We write the primal program as follows:

$$\sup_\mu \int_{\mathbb{R}^k} f(x)\mu(x)dx$$
$$\int_{\mathbb{R}^k} x(I)\mu(x)dx = \sigma_I, \quad \forall I \in I(k, n), \tag{3.1}$$
$$\int_{\mathbb{R}^k} \mu(x) = 1.$$

The supremum is over all probability measures $\mu$ on $\mathbb{R}^k$. As in the finite dimensional case, feasible solutions to the dual program correspond to degree $k$ approximating polynomials. The dual can be written as

$$\inf_{a \in \mathbb{R}^{I(k,n)}} \sum_{I \in I(k,n)} a_I \sigma_I \tag{3.2}$$

$$\sum_I a_I x(I) \geq f(x), \quad \forall x \in \mathbb{R}^k. \tag{3.3}$$

The issue here is that in general, strong duality does not hold for semi-infinite linear programs. In our case, however, where the $\sigma_i$'s are obtained as moments from a distribution $\mathcal{D}$ (as opposed to just arbitrary reals), it turns out that strong duality does hold. To see this, we note that the above primal LP is a special case of the so-called *generalized moment problem* LP, a classical problem from probability and analysis that asks if there exists a multivariate distribution with moments specified by the $\sigma_i$'s. In our case, feasibility is immediate, as the $\sigma_i$'s are obtained from $\mathcal{D}$.

As for strong duality, it is known that if the $\sigma_i$'s are in the interior of a particular set (the details are not relevant here), then the optimal value of the primal equals the optimal value of the dual. In the case that the $\sigma_i$'s do not satisfy this condition, strong duality holds assuming we relax the dual program constraints to some subset $\Omega \subseteq \mathbb{R}^n$. One concern is that we will now obtain an optimal approximating polynomial with respect to some distribution $\mathcal{D}'$ defined on $\Omega$ (as opposed to the original $\mathcal{D}$). But it is also known that in this case, *all* feasible distributions are supported on $\Omega$. As such, approximation with respect to $\mathcal{D}'$ is equivalent to approximation with respect to $\mathcal{D}$. We refer the reader to Bertsimas and Popescu [BP] (Section 2) for more details and references.

We start with an important definition.

**Definition 3.2.** *Given two distributions $\mathcal{D}, \mathcal{D}'$ on $\mathbb{R}^n$, $k \geq 0$, we say $\mathcal{D}'$ $k$ moment-matches $\mathcal{D}$ if for all $I \in I(k, n)$, $\mathbb{E}_{X \leftarrow \mathcal{D}}[X(I)] = \mathbb{E}_{X \leftarrow \mathcal{D}'}[X(I)]$.*

We can now prove the main lemma of this section:

**Lemma 3.3.** *Let $f : \mathbb{R}^n \to \{0, 1\}$ and let $\mathcal{D}$ be a distribution over $\mathbb{R}^n$ with all moments finite such that the following holds: For every distribution $\mathcal{D}'$ that $k$ moment-matches $\mathcal{D}$, $|\mathbb{E}_{X \leftarrow \mathcal{D}}[f(X)] - \mathbb{E}_{X \leftarrow \mathcal{D}'}[f(X)]| < \varepsilon$. Then, there exist degree at most $k$ polynomials $P_\ell, P_u : \mathbb{R}^n \to \mathbb{R}$ such that*

- *For every $x \in Support(\mathcal{D})$, $P_\ell(x) \leq f(x) \leq P_u(x)$.*

- *For $X \leftarrow \mathcal{D}$, $\mathbb{E}[P_u(X)] - \mathbb{E}[f(X)] \leq \varepsilon$ and $\mathbb{E}[f(X)] - \mathbb{E}[P_\ell(X)] \leq \varepsilon$.*

*Proof.* Let $opt^*$ be the value of the primal program Equation 3.1. Then, by hypothesis $opt^* < \gamma + \varepsilon$, where $\gamma = \mathbb{E}_{X \leftarrow \mathcal{D}}[f(X)]$.

Now, from the above discussion, strong duality (almost) holds for the programs in Equations (3.1) and (3.2), and we conclude that there exists a dual solution $a \in \mathbb{R}^{I(k,n)}$ with value exactly $opt^*$ that satisfies the inequality constraints for all $x \in Support(\mathcal{D})$. Define,

$$P_u(x_1, \ldots, x_n) = \sum_{I \in I(k,n)} a_I x(I).$$

Then, $P_u(\ )$ is a degree at most $k$ polynomial, and $P_u(x) \geq f(x)$ for every $x \in Support(\mathcal{D})$. Further, the assumption in the lemma implies $\mathbb{E}_{X \leftarrow \mathcal{D}}[P_u(X)] = \sum_{I \in I(k,n)} a_I \sigma_I = opt^* < \gamma + \varepsilon$. We have the existence of the lower sandwiching polynomial $P_\ell$ similarly. □

## 3.2 The Classical Moment Problem

In the previous section, we reduced the problem of constructing low-degree sandwiching polynomial approximators with respect to $\mathcal{D}$ to understanding the optimal value of a semi-infinite linear program. The feasible solutions of the linear program correspond to all distributions that are $k$ moment-matching to $\mathcal{D}$. As such, for any $k$ moment-matching distribution $\mathcal{D}'$ we need to bound $|E_\mathcal{D}[f] - E_{\mathcal{D}'}[f]|$. In this section, we give some techincal results that help us bound this difference provided the moments of $\mathcal{D}$ do not grow too fast.

We begin with the following result showing that multivariate distributions whose marginals have matching lower order moments have close characteristic functions (as quantified by $\lambda$-metric) provided the moments are well behaved.

**Theorem 3.4** (Theorem 2, Page 171, [KR]). *Let $X, Y \in \mathbb{R}^m$ be two random variables such that for any $t \in \mathbb{R}^m$, the real-valued random variables $\langle t, X \rangle, \langle t, Y \rangle$ have identical first $2k$ moments. Then, for a universal constant $C$,*

$$\mathsf{d}_\lambda(X, Y) \leq C\beta_k^{-1/4}\left(1 + \mu_2(X)^{1/2}\right),$$

*where $\mu_j(X) = \sup\{\mathbb{E}[|\langle t, x\rangle|^j] : t \in \mathbb{R}^m, \|t\| \leq 1\}$, and $\beta_k = \beta_k(X) = \sum_{j=1}^k 1/\mu_{2j}(X)^{1/2j}$.*

We now need to convert the above bound on closeness of characteristic functions to more direct measures of closeness like Levy or Kolmogorov-Smirnov metrics. Such inequalities play an important role in Fourier theoretic proofs of limit theorems (eg., Esseen's inequality; cf. Chapter XVI [Fel]) and here we use the following multi-dimensional version due to Gabovich [Gab].

**Theorem 3.5** ([Gab] Equation (8)). *Let $X, Y \in \mathbb{R}^m$ be two vector-valued random variables. Then, for a universal constant $C$ and all sufficiently large $N, T > 0$,*

$$\mathsf{d}_{\mathsf{LV}}(X, Y) \leq \int_{\frac{1}{NT} \leq t_1,\ldots,t_m \leq T} \frac{C|\varphi_X((t_1,\ldots,t_m)) - \varphi_Y(t_1,\ldots,t_m)|}{t_1 t_2 \cdots t_m} dt_1 \cdots dt_m +$$
$$\frac{C(\log T)(\log(NT))}{T} + \Pr[X \notin [-N, N]^m] + \Pr[Y \notin [-N, N]^m].$$

The above theorem leads to the following concrete relation between $\mathsf{d}_\lambda$ and $\mathsf{d}_{\mathsf{LV}}$.

**Lemma 3.6.** *Let $X, Y$ be two vector-valued random variables with $\mathsf{d}_\lambda(X, Y) \leq \delta$. Let $N(\varepsilon) \in \mathbb{R}$ be such that $\Pr[X \notin [-N(\varepsilon), N(\varepsilon)]^m], \Pr[Y \notin [-N(\varepsilon), N(\varepsilon)]^m] \leq \delta$. Then,*

$$\mathsf{d}_{\mathsf{LV}}(X, Y) \leq O\left((\log N(\delta) + 2\log(1/\delta))^m \cdot \delta\right).$$

*Proof.* Without loss of generality suppose that $\delta < 1/m^2$, as else the statement is trivial. Let $T^*$ be the value of $T$ that attains the minimum in the definition of $\mathsf{d}_\lambda$:

$$\mathsf{d}_\lambda(X, Y) = \max\{\max_{\|t\| \leq T^*}\{|\varphi_X(t) - \varphi_Y(t)|\}, 1/T^*\}.$$

As $\mathsf{d}_\lambda(X, Y) \leq \delta$, $T^* \geq 1/\delta$. Therefore, for every $t \in \mathbb{R}^m$ with $\|t\| \leq 1/\delta$, $|\varphi_X(t) - \varphi_Y(t)| \leq \delta$. Thus, applying Theorem 3.5 with $N = N(\delta)$ and $T = 1/\delta\sqrt{m}$, we get

$$\mathsf{d}_{\mathsf{LV}}(X, Y) \leq C \int_{\frac{1}{NT} \leq t_1,\ldots,t_m \leq T} \frac{|\varphi_X(t) - \varphi_Y(t)|}{t_1 \cdots t_m} dt + O(\log^2(NT) \cdot \delta\sqrt{m}) + O(\delta)$$
$$\leq C \int_{\frac{1}{NT} \leq t_1,\ldots,t_m \leq T} \frac{\delta}{t_1 \cdots t_m} dt + O(\log^2(NT) \cdot \delta\sqrt{m})$$
$$\leq C\delta \cdot (\log N + 2\log T)^m + O(\log^2(NT) \cdot \delta\sqrt{m})$$
$$= O((\log N(\delta) + 2\log(1/\delta))^m \cdot \delta).$$

□

## 3.3 Low-Order Moments, Functions of Halfspaces, and Log-Concave Densities

We are now ready to complete the proof of the main theorem for learning functions of halfspaces with respect to log-concave distributions - Theorem 1.1. We do so by using the tools from the previous section on moment bounds to analyze the optimum value of the primal LP from Section 3. This will imply low-degree $\ell_1$ approximators with low error for any $f \in \mathcal{C}$. We can then apply known results due to Kalai et al. [KKMS] (Theorem 2.2) relating approximability by low-degree polynomials and agnostic learning.

*Proof of Theorem 3.1.* Without loss of generality suppose that $\mathcal{D}$ is in isotropic position. We can do so, as any distribution can be brought to isotropic position by an affine transformation and the class of intersections of halfspaces is invariant under affine transformations.

Let halfspace $h_i : \mathbb{R}^n \to \{1, -1\}$ be given as $h_i(x) = \text{sign}(\langle w_i, x \rangle - \theta_i)$ for $w_i \in \mathbb{R}^n$ with $\|w_i\| = 1$ and $\theta_i \in \mathbb{R}$.

Let $X \leftarrow \mathcal{D}$ and let $X' \leftarrow \mathcal{D}'$, where $\mathcal{D}'$ is any distribution that is $2k$-moment matching to $\mathcal{D}$ for $k = 2^{O((m/\varepsilon)^4)}$ to be chosen later. Let $Y = (\langle w_1, X \rangle, \langle w_2, X \rangle, \cdots, \langle w_m, X \rangle)$ and $Y' = (\langle w_1, X' \rangle, \cdots, \langle w_m, X' \rangle)$. Observe that for every $t \in \mathbb{R}^m$, the first $2k$ moments of $\langle t, Y \rangle, \langle t, Y' \rangle$ are identical. Thus, we can apply Theorem 3.4 to the random variables $Y, Y'$. For $t \in \mathbb{R}^m$, $\|t\| = 1, j > 0$,

$$\mathbb{E}[|\langle t, Y \rangle|^j] = \mathbb{E}[|\langle \sum_{r=1}^m t_r w_r, X \rangle|^j] \leq j^j \cdot \mathbb{E}[\langle \sum_{r=1}^m t_r w_r, X \rangle^2]^{j/2} = j^j \cdot \|\sum_{r=1}^m t_r w_r\|^j \leq j^j m^j,$$

where the first inequality follows from Theorem 2.3 and the second equality from $X$ being isotropic. Therefore, for $\mu_j(Y)$ and $\beta_k$ as defined in Theorem 3.4,

$$\beta_k = \sum_{j=1}^k \frac{1}{\mu_j(Y)^{1/2j}} \geq \sum_{j=1}^k \frac{1}{2mj} = \Omega((\log k)/m). \quad (3.4)$$

We now wish to get a good estimate on $N(\delta)$ as defined in Lemma 3.6. From Theorem 2.3 and Markov's inequality, for every $\alpha > 0$, and $r \in [m], j \leq 2k$ even

$$\Pr[|\langle w_r, X \rangle| > \alpha] \leq \frac{\mathbb{E}[\langle w_r, X \rangle^j]}{\alpha^j} \leq \frac{j^j}{\alpha^j}.$$

Therefore, for $j = \log(m/\delta)$, and $\alpha = 2j$, $\Pr[|\langle w_r, X \rangle| > 2j] < \delta/m$. Thus, by using a union bound over all the components of $Y$, for $N = 2j = 2\log(m/\delta)$,

$$\Pr[Y \notin [-N, N]^m] < \delta. \quad (3.5)$$

As the above calculation only involved the first $2k$ moments of $X$, the same property should hold for $Y'$. From Equations (3.4), (3.5) and Theorem 3.4,

$$\mathsf{d}_\lambda(Y, Y') \leq O\left(\frac{m^{1/4}}{\log^{1/4} k}\right). \quad (3.6)$$

Let $k = 2^{O(m/\delta^4)}$ be large enough so that the above error bound is $\mathsf{d}_\lambda(Y, Y') \leq \delta$. Therefore, from Lemma 3.6,
$$\mathsf{d}_{\mathsf{LV}}(Y, Y') \leq (\log((\log m)/\delta))^{O(m)} \cdot \delta.$$

Now observe that by Theorem 2.4, for every $r \in [m]$, $t \in \mathbb{R}$, $\alpha > 0$, $\Pr[Y_r \in [t, t+\alpha]] = O(\alpha)$. Thus, from the above equation and Fact 2.5,
$$\mathsf{d}_{\mathsf{cdf}}(Y, Y') \leq O(m \cdot \mathsf{d}_{\mathsf{LV}}(Y, Y')) = (\log((\log m)/\delta))^{O(m)} \cdot \delta \equiv \varepsilon. \tag{3.7}$$

Since the above argument worked for any weight vectors $w_1, \ldots, w_m \in \mathbb{R}^m$, a similar argument applied to weight vectors $a_1 w_1, a_2 w_2, \ldots, a_m w_m$ for $a \in \{1, -1\}^m$, gives
$$\mathsf{d}_{\mathsf{cdf}}((a_1 Y_1, \ldots, a_m Y_m), (a_1 Y_1', \ldots, a_m Y_m')) \leq \varepsilon.$$

Therefore, by Lemma 2.6 applied to $Y, Y'$ and $g$, $|\mathbb{E}[f(X)] - \mathbb{E}[f(X')]| \leq 2^m \varepsilon$.

Hence, by Lemma 3.3, for $P \equiv P_u$ a degree at most $k$ polynomial as in Lemma 3.3,
$$\mathbb{E}[|P(X) - f(X)|] = \mathbb{E}[P(X)] - \mathbb{E}[f(X)] \leq 2^m \varepsilon.$$

The theorem now follows from setting $\varepsilon = \varepsilon'/2^m$ as $k = 2^{O(m/\delta^4)} = 2^{(\log((\log m)/\varepsilon)^{O(m)})/\varepsilon^4}$. □

## 4 Smoothed Complexity of Learning Functions of Halfspaces

We now consider the smoothed complexity of learning convex sets defined by intersections of halfspaces and extend our learning results to handle any distribution whose marginals obey a subexponential tail bound. We feel this is a mild restriction to place on the distribution. It is well known that any (isotropic) log-concave distribution obeys such a tail bound.

Our high level approach will be similar to that for log-concave densities: we use moment bounds and results from Section 3.2 to show that functions of halfspaces cannot distinguish (smoothed) distributions with strong moment bounds. Adding a Gaussian perturbation plays an important role in our setting, by essentially allowing us to impose certain *probabilistic* margin constraints in the form of anti-concentration bounds. One interpretation of our results is that in the setting of smoothed-analysis, learning geometric classes becomes easier in many cases because the underlying Gaussian perturbation makes the distribution anti-concentrated (i.e., no sharp peaks) "for free."

We first state our results and defer the proofs to the following sections.

**Theorem 4.1.** *Let $\mathcal{D}$ be a sub-exponential distribution over $\mathbb{R}^n$. Let $h_1, \ldots, h_m : \mathbb{R}^n \to \{1, -1\}$, be halfspaces and let $g : \{1, -1\}^m \to \{1, -1\}$ be an arbitrary function. Define $f : \mathbb{R}^n \to \{1, -1\}$ by $f(x) = g((h_1(x), \ldots, h_m(x)))$. Then, for every $\sigma > 0$, there exists a polynomial $P$ of degree at most*
$$k = \exp((\log((\log m)/\sigma\varepsilon))^{O(m)}/(\sigma\varepsilon)^4)$$
*such that $\mathbb{E}_{X \leftarrow \mathcal{D}(\sigma)}[|f(X) - P(X)|] \leq \varepsilon$.*

**Theorem 4.2.** *Let $\mathcal{D}$ be a sub-Gaussian distribution over $\mathbb{R}^n$. Let $h_1, \ldots, h_m : \mathbb{R}^n \to \{1, -1\}$, be halfspaces and let $g : \{1, -1\}^m \to \{1, -1\}$ be an arbitrary function. Define $f : \mathbb{R}^n \to \{1, -1\}$ by $f(x) = g((h_1(x), \ldots, h_m(x)))$. Then, for every $\sigma > 0$, there exists a real-valued polynomial $P$ of degree at most $k = (\log((\log m)/\sigma\varepsilon))^{O(m)}/(\sigma\varepsilon)^4$ such that $\mathbb{E}_{X \leftarrow \mathcal{D}(\sigma)}[|f(X) - P(X)|] \leq \varepsilon$.*

Theorem 1.3 and the precise runtimes as given in Table 1.2 follow from the above results and Theorem 2.2.

## 4.1 Sub-Exponential Densities

In this section we study sub-exponential densities and prove Theorem 4.1.

**Definition 4.3.** *We say an isotropic distribution $\mathcal{D}$ on $\mathbb{R}^n$ is* sub-exponential *if there exist constants $C, \alpha > 0$, such that for every $w \in \mathbb{R}^n$, $\|w\| = 1$, and $t > 0$,*

$$\Pr_{X \leftarrow \mathcal{D}}[|\langle w, X \rangle| > t] < C \exp(-\alpha t).$$

*More generally, we say a distribution $\mathcal{D}$ on $\mathbb{R}^n$ is sub-exponential if the isotropic distribution obtained by putting $\mathcal{D}$ in an isotropic position by an affine transformation is sub-exponential.*

We shall use the following standard fact giving strong moment bounds for random variables with sub-exponential tails.

**Fact 4.4.** *Let $X$ be unit variance random variable such that $\Pr[|X| > t] < C \exp(-\alpha t)$. Then, for all $k > 0$, $\mathbb{E}[|X|^k] < C(k/\alpha)^k$.*

Finally, we need the following fact showing that convolving any distribution with a Gaussian distribution leads to *anti-concentration*.

**Fact 4.5.** *For any real-valued random variable $X$, $Z \leftarrow \mathcal{N}(0, \sigma)$ and $t \in \mathbb{R}$, $\alpha > 0$, $\Pr[X + Z \in [t, t+\alpha]] < C\alpha/\sigma$, where $C$ is a universal constant.*

*Proof.* Fix $t \in \mathbb{R}$ and $\alpha > 0$. Then,

$$\Pr[Z \in [t, t+\alpha]] = \int_t^{t+\alpha} \frac{e^{-\frac{x^2}{2\sigma^2}}}{\sqrt{2\pi\sigma^2}} dx < \int_t^{t+\alpha} \frac{1}{\sqrt{2\pi\sigma^2}} dx = C\alpha/\sigma,$$

where $C = 1/\sqrt{2\pi}$. The claim now follows from:

$$\Pr[X + Z \in [t, t+\alpha]] = \mathbb{E}_X[\Pr_Z[Z \in [(t-X), (t-X)+\alpha]]] < \mathbb{E}_X[C\alpha/\sigma] < C\alpha/\sigma.$$

□

*Proof of Theorem 4.1.* The proof follows the same approach as that of Theorem 3.1. Without loss of generality, we can suppose that $\mathcal{D}$ is in isotropic position as functions of halfspaces are closed under affine transformations.

Let the Gaussian perturbation be $Z \leftarrow \mathcal{N}(0, \Sigma)^m$, where $\Sigma \succeq \sigma \mathsf{I}_m$. We next renormalize the distribution $\mathcal{D}$ so that $\mathcal{D}(\sigma)$ is in isotropic position. Note that $\mathcal{D}(\sigma)$ is also sub-exponential. This follows from a simple union bound. For any direction $w \in \mathbb{R}^m$, $\|w\| = 1$, and $X \leftarrow \mathcal{D}$ and $Z \leftarrow \mathcal{N}(0, \Sigma)^m$,

$$\Pr[|\langle X+Z, w \rangle| > t] \leq \Pr[|\langle X, w \rangle| > t/2] + \Pr[|\langle Z, w \rangle| > t/2] = O(\exp(-\Omega(t))),$$

where the last inequality follows from the fact that $X$ is sub-exponential by definition and that the uni-variate Gaussian distribution is sub-exponential.

Fix halfspaces $h_i : \mathbb{R}^n \to \{1, -1\}$ and let random variables $X \leftarrow \mathcal{D}(\sigma)$ and let $X' \leftarrow \mathcal{D}'$, where $\mathcal{D}'$ $k$-moment matches $\mathcal{D}$ for $k$ to be chosen later. Let $Y, Y'$ be as in Theorem 3.1. Then, by Fact 4.4, for any $w \in \mathbb{R}^n$, $\|w\| = 1$, $\mathbb{E}[|\langle w, X \rangle|^j] < C(j/\alpha)^j$.

Observe that the proofs of Equations (3.4) and (3.5) in Theorem 3.1 only used moment bounds for log-concave distributions, and sub-exponential distributions have similar bounds on moments. Thus, by similar arguments, for $k = 2^{O(m/\delta^4)}$ sufficiently large, we get

$$\mu_j(Y) \leq C(jm/\alpha)^j, \quad \beta_k(Y) = \Omega((\log k)/m), \tag{4.1}$$

and for $N = O(\log(m/\delta)/\alpha)$ sufficiently large,

$$\Pr[Y \notin [-N, N]^m] + \Pr[Y' \notin [-N, N]^m] < 2\delta.$$

Combining the above two equations and Lemma 3.6, we have

$$\mathsf{d}_\lambda(Y, Y') = (\log(\log(m/\delta)))^{O(m)} \cdot \delta. \tag{4.2}$$

Now, note that for any $r \in [m]$, $Y_r$ can be written as $Y'_r + Z_r$, where $Z_r \leftarrow \mathcal{N}(0, \sigma)$ is independent of $Y'_r$. Therefore, by Fact 4.5, $\Pr[Y_r \in [t, t+\gamma]] = O(\gamma/\sigma)$ for $t \in \mathbb{R}, \alpha > 0$. Thus, by the above equation and Fact 2.5,

$$\mathsf{d}_{\mathsf{cdf}}(Y, Y') = O(m\mathsf{d}_\lambda(Y, Y')/\sigma) = (\log(\log(m/\delta)))^{O(m)} \cdot \delta/\sigma = \varepsilon.$$

The theorem now follows from an argument similar to that of Theorem 3.1 following Equation 3.7. □

## 4.2 Sub-Gaussian Densities

We now study sub-Gaussian densities and show an analogue of Theorem 4.1 with much better parameters. The improvement in parameters comes from the fact that sub-Gaussian have much more tightly controlled moments.

**Definition 4.6.** *We say an isotropic distribution $\mathcal{D}$ is sub-Gaussian if there exist constants $C, \alpha > 0$, such that for every $w \in \mathbb{R}^n$, $\|w\| = 1$, and $t \in \mathbb{R}$,*

$$\Pr_{X \leftarrow \mathcal{D}}[|\langle w, X \rangle| > t] < C \exp(-\alpha t^2).$$

*More generally, we say a distribution $\mathcal{D}$ on $\mathbb{R}^n$ is sub-exponential if the isotropic distribution obtained by putting $\mathcal{D}$ in an isotropic position by an affine transformation is sub-exponential.*

Analogous to Fact 4.4, we have the following statement for sub-gaussian densities.

**Fact 4.7.** *Let $X$ be unit variance random variable such that $\Pr[|X| > t] < C \exp(-\alpha t)$. Then, $\mathbb{E}[|X|^k] < C(k/\alpha^2)^{k/2}$.*

*of Theorem 4.2.* The proof follows the same approach as that of Theorem 4.1. We only highlight the important differences. Fix halfspaces $h_i : \mathbb{R}^n \to \{1, -1\}$, and random variables $X, X', Y, Y'$ as in the proof of Theorem 4.1. Now, observe that for $k = \Omega(m/\delta^4)$, sufficiently large, for any $t \in \mathbb{R}^m, \|t\| = 1$, and $j > 0$,

$$\mathbb{E}[|\langle t, Y\rangle|^j] = \mathbb{E}[|\langle \sum_{r=1}^m t_r w_r, X\rangle|^j]$$
$$\leq Cj^{j/2} \cdot \mathbb{E}[\langle \sum_{r=1}^m t_r w_r, X\rangle^2]^{j/2}/\alpha^j \quad \text{(Fact 4.7)}$$
$$= O((m/\alpha)^j \cdot j^{j/2}).$$

Therefore,

$$\beta_k(Y) = \sum_{j=1}^k \frac{1}{\mu_{2j}(Y)^{1/2j}} \geq \sum_{j=1}^k \frac{\alpha}{m\sqrt{2j}} = \Omega(\sqrt{k}/m). \quad (4.3)$$

Note that the above bound on $\beta_k$ is exponentially better than the $\Omega(\log k)$ bound we had for log-concave and sub-exponential densities and this leads to the quantitative improvements for sub-Gaussian densities.

Now, by using Markov's inequality it follows that for $k > \log(m/\delta)$, and $N = O(\sqrt{\log(m/\delta)/\alpha})$ sufficiently large,
$$\Pr[Y \notin [-N, N]^m] + \Pr[Y' \notin [-N, N]^m] < 2\delta.$$

Combining the above two equations and Lemma 3.6, we get

$$\mathsf{d}_\lambda(Y, Y') = (\log(\log(m/\delta)))^{O(m)} \cdot \delta.$$

The theorem now follows from the above inequality and an argument similar to that of Theorem 4.1 following Equation 4.2. $\square$

### 4.3 Non-Product Distributions on Hypercube

Learning intersections of halfspaces with respect to distributions on the hypercube is a long-standing and fundamental open problem in learning theory. To date, most non-trivial results pertain to product distributons on the hypercube, with the exception of the work of Wimmer [Wim] who can handle symmetric distributions on the hypercube.

Our results imply algorithms for agnostically learning functions of halfspces in the smoothed complexity setting for distributions on the hypercube that are locally-independent. Specifically, call a distribution $\mathcal{D}$ on $\{1, -1\}^n$ $k$-wise independent if for any $I \subseteq [n], |I| \leq k, X \leftarrow \mathcal{D}$, the variables $(X_i : i \in I)$ are independent. (This is the same as saying $\mathcal{D}$ $k$-moment matches the uniform distribution on $\{1, -1\}^n$). Our learning algorithms for sub-Gaussian densities, Theorem 4.2, immediately imply the following for learning with respect to $k$-wise independent distributions.

**Theorem 4.8.** *For all $m, \varepsilon, \sigma$ there exists $k = O_{m,\varepsilon,\sigma}(1)$ such that the following holds. Functions of $m$ halfspaces are agnostically learnable with respect to any $k$-wise independent distribution on $\{1, -1\}^n$ in time $n^{O_{m,\varepsilon,\sigma}(1)}$ where $\varepsilon$ is the accuracy and $\sigma$ is the perturbation parameter.*

In contrast, it is not clear if any of the previous techniques can give algorithms for learning intersections of halfspaces that are even $\Omega(n)$-wise independent.

*Proof.* The uniform distribution on $\{1,-1\}^n$ is known to be sub-Gaussian [Pin]. Further, observe that in the proof of Theorem 4.2 we only used properties of the first $k$-moments for $k = (\log((\log m)/\sigma\varepsilon))^{O(m)}/(\sigma\varepsilon)^4$. Thus, the same arguments should work for any distribution $\mathcal{D}$ which is $k$-wise independent. The thoerem then follows from combining the direct analogue of Theorem 4.2 for $k$-wise independent distributions $\mathcal{D}$ with Theorem 2.2. □

## 5 Bounded Independence Fools Degree Two Threshold Functions

Here we show that the methods of Section 3 can also be used with respect to the uniform distribution over $\{1,-1\}^n$. We use the moment-matching techniques to give a new proof for the recent result of Diakonikolas, Kane, and Nelson [DKN] that bounded independence fools degree-2 polynomial threshold functions. Our proof gives worse parameters, but is considerably different and is perhaps simpler. We also establish a connection between the pseudorandomness problem and the well studied classical moment problem in probability (see [Akh] for instance).

**Theorem 5.1.** *There exist constants $C, C'$ such that the following holds. Let $\mathcal{D}$ be a $m$-wise independent distribution over $\{1,-1\}^n$ for $m = 2^{C/\delta^9}$. Then, for every degree $2$ polynomial $P : \mathbb{R}^n \to \mathbb{R}$, and $x \leftarrow \mathcal{D}$, $y \in_u \{1,-1\}^n$, $\mathsf{d}_{\mathsf{cdf}}(P(x), P(y)) < C'\delta$. In other words, $(2^{\tilde{O}(1/\delta^9)})$-wise independence $\delta$-fools degree two threshold functions.*

In comparison, Diakonikolas et al. show that $\tilde{O}(\delta^{-9})$-wise independence suffices. This bound was later improved to $O(\delta^{-8})$ in [Kan2].

We shall use the following quantitative estimate due to Klebanov and Mkrtchyan which can be seen as a one dimensional version of Theorem 3.4, albeit with better parameters.

**Theorem 5.2** (Theorem 1, [KM1])**.** *Let $X, Y$ be real-valued random variables with $\mathbb{E}[X^i] = \mathbb{E}[Y^i]$ for $1 \leq i \leq 2m$ and $\mathbb{E}[X^2] = 1$. Then, for a universal constant $C > 0$,*

$$\mathsf{d}_{\mathsf{LV}}(X, Y) \leq \frac{C_\sigma \cdot \ln(1 + \beta_m(X))}{\beta_m(X)^{1/4}}.$$

We only detail the case of *regular* polynomials here, the reduction from the general case to the regular case works via the regularity lemma of Harsha et al., [HKM] and Diakonikolas et al., [DSTW].

**Definition 5.3.** *A multi-linear polynomial $P : \mathbb{R}^n \to \mathbb{R}$, $P(x) = \sum_{I \subseteq [n]} a_I \prod_{i \in I} x_i$ is $\delta$-regular if for every $i \in [n]$,*

$$\sum_{i=1}^n \left( \sum_{I \subseteq [n], I \ni i} a_I^2 \right)^2 \leq \delta^2 \|P\|_2^4,$$

*where $\|P\|_2^2 = \sum_I a_I^2$.*

**Theorem 5.4.** *There exist constants $C, C'$ such that the following holds. Let $\mathcal{D}$ be a $m$-wise independent distribution over $\{1,-1\}^n$ for $m = 2^{C/\delta^2}$. Then, for every $\delta$-regular degree $2$ polynomial $P : \mathbb{R}^n \to \mathbb{R}$, and $x \leftarrow \mathcal{D}$, $y \in_u \{1,-1\}^n$, $\mathsf{d}_{\mathsf{cdf}}(P(x), P(y)) < C'\delta^{2/9}$.*

Theorem 5.1 follows from the above theorem and the regularity lemma of Harsha et al., Diakonikolas et al. We refer the reader to the work of Meka and Zuckerman [MZ] for a similar reduction of the general case to the regular case in the pseudorandomness context and omit it here.

To prove Theorem 5.4 we use the following results about low-degree polynomials. The lemma gives us control on how fast the moments of low-degree polynomials grow.

**Theorem 5.5** (Hypercontractivity, [LT]). *For $1 < p < q < \infty$, and $P : \mathbb{R}^n \to \mathbb{R}$ a degree $d$ polynomial, the following holds:*

$$\mathop{\mathbb{E}}_{X \in_u \{1,-1\}^n}[|P(X)|^q]^{1/q} \leq \left(\frac{q-1}{p-1}\right)^{d/2} \mathop{\mathbb{E}}_{X \in_u \{1,-1\}^n}[|P(X)|^p]^{1/p}. \tag{5.1}$$

The next two theorems helps us get anti-concentration bounds for regular polynomials over the hypercube.

**Theorem 5.6** (Mossel et al. [MOO]). *There exists a universal constant $C$ such that the following holds. Let $P : \mathbb{R}^n \to \mathbb{R}$ be a degree $d$ $\delta$-regular (multi-linear) polynomial. Then, for $x \in_u \{1,-1\}^n$ and $y \leftarrow \mathcal{N}(0,1)^n$,*

$$\mathsf{d}_{\mathsf{cdf}}(P(x), P(y)) \leq C\, d\, \delta^{2/(4d+1)}.$$

**Theorem 5.7** (Carbery and Wright [CW]). *There exists an absolute constant $C$ such that for any polynomial $Q$ of degree at most $d$ with $\|Q\| = 1$ and any interval $I \subseteq \mathbb{R}$ of length $\alpha$, $\Pr_{X \leftarrow \mathcal{N}(0,1)^n}[Q(X) \in I] \leq C d\, \alpha^{1/d}$.*

*of Corollary 5.4.* It suffices to show the statement when $\mathcal{D}$ is $4m$-wise independent for $m = 2^{C/\delta^2}$ for $C$ to be chosen later. Without loss of generality suppose that $\|P\| = 1$. Let random variables $X = P(x)$, for $x \leftarrow \mathcal{D}$ and $Y = P(y)$, for $y \in_u \{1,-1\}^n$. Then, $\mathbb{E}[X^i] = \mathbb{E}[Y^i]$ for $i \leq 2m$ as $x$ is $4m$-wise independent and $P$ is a degree 2 polynomial. Now, for $i \leq m$, by hypercontractivity, Theorem 5.5, applied to $q = i$, $d = 2$,

$$\mathbb{E}[X^{2i}] = \mathbb{E}[Y^{2i}] < (2i)^{2i}.$$

Therefore,

$$\beta_m = \sum_{i=1}^m \frac{1}{\mathbb{E}[X^{2i}]^{1/2i}} > \sum_{i=1}^m \frac{1}{2i} = \Omega(\log m).$$

By Theorem 5.2,

$$\mathsf{d}_{\mathsf{LV}}(X, Y) = O\left(\frac{\log\log m}{(\log m)^{1/4}}\right).$$

Now, by Theorem 5.6 and Theorem 5.7 applied to degree $d = 2$, $\sup_t \Pr[Y \in [t, t+\alpha]] = O(\delta^{2/9} + \sqrt{\alpha})$. Therefore, by Fact 2.5

$$\mathsf{d}_{\mathsf{cdf}}(X, Y) = O\left(\delta^{2/9} + \frac{\sqrt{\log\log m}}{(\log m)^{1/8}}\right).$$

The statement now follows by choosing $C$ to be sufficiently large. □